\documentclass{elsart}
\usepackage{amssymb}

\usepackage{graphicx}

\journal{Physics Letters A}
\begin{document}

\begin{frontmatter}

\title{Quantum pathology of static internal imperfections in flawed quantum computers}
\author{Murat \c{C}etinba\c{s}\corauthref{cor}},
\corauth[cor]{Corresponding author.} 
\ead{cetinbas@sfu.ca}
\author{Joshua Wilkie}
\ead{wilkie@sfu.ca}
\address{Department of Chemistry, Simon Fraser University, Burnaby, British Columbia V5A 1S6, Canada}

\begin{abstract}
Even in the absence of external influences the operability of a quantum 
computer (QC) is not guaranteed because of the effects 
of residual one-- and two--body imperfections. Here we investigate how 
these internal flaws affect the performance of a
quantum controlled-NOT (CNOT) gate in an isolated flawed QC. First we 
find that the performance of the CNOT gate 
is considerably better when the two--body imperfections are strong. Secondly, we find
that the largest source of error is due to a coherent shift rather than decoherence or dissipation. Our results suggest
that the problem of internal imperfections should be given much more attention in designing scalable QC
architectures.
\end{abstract}

\end{frontmatter}
\section{Introduction}
Quantum computers (QCs) are subject to internal as well as external errors. External errors arising from system--environment interactions (i.e. decoherence and dissipation \cite{DC}) are perceived to be the primary obstacle to practical implementation of quantum computation\cite{Book}. However, even in the absence of a macroscopic environment the efficient operability of a QC is not guaranteed. Unavoidable one-- and two--body static internal imperfections can also be potentially harmful error sources. 

Recently, the problem of internal errors 
has attracted some attention. Studies on the
statistical properties of eigenspectra of flawed QCs showed that sufficiently strong 
residual two--qubit interactions cause the emergence 
of chaos\cite{GS,GS2} and a consequent thermalization of the QC core\cite{BCS} which is believed to be a serious obstacle for quantum computation. Destructive effects of chaos on QC operation
include an incoherent mixing of the ideal (i.e. separable) computational states of the qubits\cite{GS}. This mixing is expected to worsen with time resulting in an effective QC meltdown\cite{GS2}. However, dynamical calculations in which the effects of internal 
flaws were modeled by random kicks showed better fidelity\cite{Prosen,Frahm} decay for chaotic than for
integrable perturbations. These seemingly contradictory conclusions, i.e. destruction and stabilization of QCs with chaos, suggest that determining the effects of internal flaws is a problem of considerable complexity. 

The pathology of internal errors includes internal decoherence and dissipation
and even coherent shifting of the QC operation. The relative importance of these effects
and their dependence on residual internal couplings can only be determined via
fully realistic dynamical simulations of QC operation. The objective of this Letter is to directly
examine flawed QC operation in a specific architecture, to investigate the different types of internal errors that emerge, and to observe how these errors change with varying magnitudes of internal flaws. 

A suggested remedy for the errors caused by these flaws is to operate the QC below the chaos border. This could be 
achieved by employing weaker coupling. This can be a problematic solution\cite{GS2}, however. First of all, the duration of a gate operation is inversely proportional to gate coupling strength, and longer gate operations are more likely to be decohered due to the external influences. Secondly, even below the chaos border (i.e. in the integrable regime), the residual two-qubit interactions are still strong enough to mix ideal logic states of qubits\cite{weak}, and thus residual entanglement among qubits is unavoidable. In addition there is evidence that the irreversible effects of internal decoherence are more harmful in the integrable regime\cite{Tess,Mil}, although chaos is indeed harmful at high temperatures\cite{Sanz}. 

We focus our study on the dynamics of a two-qubit subsystem located at the center of a 3x4 qubit 2D grid circuit. The qubits
are modeled and parametrized based on a Josephson charge qubit QC\cite{Nori} proposal. This scalable architecture has an extremely long external decoherence time of $\sim 10^{-4}$ s, which in principle allows around $10^{6}$ single qubit operations. However, internal imperfections for this design could be a limiting factor. Our two-qubit subsystem (i.e. the active part) 
performs a quantum controlled-NOT (CNOT) gate while the rest of the qubits are idle. We simulate the effects of flaws by
adding residual one-- and two--body imperfections in the idle part. The ten qubits neigboring the active ones
are chosen to be idle for computational simplicity.

We monitor the dynamics of the CNOT gate by two traditional error quantifiers: purity and fidelity. Purity, while insensitive to unitary errors, gives a good measure of non-unitary errors. Fidelity, on the other hand, can detect both non-unitary and unitary errors. We find that non-unitary errors due to internal decoherence and dissipation are non-negligible even for relatively small numbers of bath qubits. However, we find that strong chaos in the bath degrees of freedom leads to complete 
suppression of non-unitary errors. Surprisingly, the system's
fidelity is much worse than its purity would suggest. The unexpectedly large differences between the purity and fidelity 
show that unitary errors generated by coherent shifting are dominant in the system dynamics. These large 
unitary errors persist even in the chaotic regime, and severely endanger the viability of quantum computation. We have 
observed similar shifts in other contexts which are discussed elsewhere\cite{CW2}.

Two important differences emerge in the modeling of external and internal errors. First, traditional boson-bath models\cite{Bbath} or spin-bath models\cite{Stamp}, in which self-interactions among bath modes are omitted, are inappropriate for 
systems with important internal bath dynamics. While some master equations\cite{SRA,Red} include these effects, they are approximate and thus cannot be totally trusted. Second, the assumption of coordinate type couplings in boson-bath models results in vanishing canonical averages for the bath coupling operator, and so no coherent shift is observed. Fortunately, the problem of internal errors in a small QC is amenable to exact propagation. We can thus avoid all the pitfalls of the approximate theories. 

The organization of this Letter as follows. In Section 2 we describe our isolated QC model and our exact numerical approach. In Section 3 we present our results for two error quantifiers, purity and fidelity, for a number of initial states and two different error generators. Section 4 discusses the observed effects in more detail.

\section{Isolated flawed QC model}

The total Hamiltonian of the isolated flawed QC reads
\begin{equation}
\hat{H}(t)=\hat{H}_{S}(t)+\hat{H}_{SB}+\hat{H}_{B}
\end{equation}
\noindent
in which $\hat{H}_{S}(t)$ is control Hamiltonian of the system (active part), $\hat{H}_{B}$ is the bath Hamiltonian (idle part), and $\hat{H}_{SB}$ is the interaction Hamiltonian. 

The system Hamiltonian to control two active charge-qubits \cite{Nori} is
\begin{equation}
\hat{H}_{S}(t)=-\frac{1}{2}\sum_{i=1}^2({\cal B}_{i}^{x}(t)\hat{\sigma}_{x}^{i} +{\cal B}_{i}^{z}(t)\hat{\sigma}_{z}^{i} )
+ {\cal J}_{x}(t)\hat{\sigma}_{x}^{1}\hat{\sigma}_{x}^{2}.
\label{ctrl}
\end{equation}
\noindent
The CNOT protocol given in Ref. \cite{Nori} can be implemented from Eq. (\ref{ctrl}) in nine steps with fast Hadamard gates, and this is summarized in Table I.
Note that ignoring all possible imperfections in (\ref{ctrl}), and assuming that consecutive gates can be simultaneously switched on and off ( i.e. no free Hamiltonian evolution is allowed) we eliminate all likely systematic unitary errors in the system. (See Ref.\cite{Errors} for a classification of errors). Hence, all unitary errors we observe will arise from interaction with the bath qubits.
 
Defining two bath interaction operators $\hat{\Sigma}_{x}=\sum_{i=3}^{N+2}\lambda_{x}^{i} \hat{\sigma}_{x}^{i}$ and 
$\hat{\Sigma}_{z}=\sum_{i=3}^{N+2} \lambda_{z}^{i} \hat{\sigma}_{z}^{i}$, the interaction Hamiltonian takes one of the two forms
\begin{equation}
\hat{H}_{SB}= ( \hat{\sigma}_{\alpha}^{1}+\hat{\sigma}_{\alpha}^{2} )\hat{\Sigma}_{\alpha}
\end{equation}
where $\alpha\in \{x,z\}$. Note that while the bit-flip errors by xx-type coupling occur naturally, we do not expect 
phase-errors by zz-type coupling to appear in this design. Nonetheless, we also explore 
phase-errors for pedagogical reasons to demonstrate how the decoherence and shift can vary for different 
system-environment interactions.

\begin{table}
\caption{The switching intervals and active Hamiltonians used to implement CNOT gate.}
\begin{center}
\begin{tabular}{ccc}
\hline 
\hline 
Switching Intervals~~~~~~~&~~~Active Hamiltonian
\tabularnewline
\hline
$[\tau_{0}=0, \tau_{1} = \pi / (2{\cal{B}}^{z})]~~~~~$
&
$-\frac{1}{2}{\cal{B}}^{z}\hat{\sigma}_{z}^{2}$
\tabularnewline
$[\tau_{1}, \tau_{2} = \tau_{1}+ \pi / (2{\cal{B}}^{x})]~~~~$
&
$-\frac{1}{2}{\cal{B}}^{x} \hat{\sigma}_{x}^{2}$
\tabularnewline
$[\tau_{2}, \tau_{3} = \tau_{2}+ \pi / (2{\cal{B}}^{z})]~~~~$
&
$+\frac{1}{2}{\cal{B}}^{z}\hat{\sigma}_{z}^{2}$
\tabularnewline
$[\tau_{3}, \tau_{4} = \tau_{3}+ \sqrt{2} \pi / (2{\cal{B}}^{z})]$
&
$-\frac{1}{2}{\cal{B}}^{z}\sum_{i=1}^{2}(\hat{\sigma}_{z}^{i}+\hat{\sigma}_{x}^{i})$
\tabularnewline
$[\tau_{4}, \tau_{5} = \tau_{4}+ \pi / (4{\cal{J}}_x)]~~~~$
&
${\cal J}_{x}(-\hat{\sigma}_{x}^{1}-\hat{\sigma}_{x}^{2}+\hat{\sigma}_{x}^{1}\hat{\sigma}_{x}^{2})$
\tabularnewline
$[\tau_{5}, \tau_{6} = \tau_{5}+ \sqrt{2} \pi / (2{\cal{B}}^{z})]$
&
$+\frac{1}{2}{\cal{B}}^{z}\sum_{i=1}^{2}(\hat{\sigma}_{z}^{i}+\hat{\sigma}_{x}^{i})$
\tabularnewline
$[\tau_{6}, \tau_{7} =\tau_{6}+ \pi / (2{\cal{B}}^{z})]~~~~$
&
$-\frac{1}{2}{\cal{B}}^{z}\hat{\sigma}_{z}^{2}$
\tabularnewline
$[\tau_{7}, \tau_{8} =\tau_{7}+ \pi / (2{\cal{B}}^{x})]~~~~$
&
$+\frac{1}{2}{\cal{B}}^{x}\hat{\sigma}_{x}^{2}$
\tabularnewline
$[\tau_{8}, \tau_{9} =\tau_{8}+ \pi / (2{\cal{B}}^{z})]~~~~$
&
$+\frac{1}{2}{\cal{B}}^{z}\hat{\sigma}_{z}^{2}$
\tabularnewline
\hline\hline
\end{tabular}\end{center}
\end{table}

Including one- and two-body static imperfections, the two-qubit Hamiltonian of the system (\ref{ctrl}) can be generalized for an N-qubit bath:
\begin{eqnarray}
\hat{H}_{B} = -\frac{1}{2} \sum_{i=3}^{N+2} \left( B_{i}^{x}\hat{\sigma}_{x}^{i} 
+ B_{i}^{z} \hat{\sigma}_{z}^{i} \right)
+ \sum_{i=3}^{N+1}\sum_{j=i+1}^{N+2} J_{x}^{i,j} \hat{\sigma}_{x}^{i}\hat{\sigma}_{x}^{j}.
\label{Hb}
\end{eqnarray} 
Here, individual weights $B_{i}^{\alpha} $ are sampled randomly and uniformly, i.e. $B_{i}^{\alpha} \in [B_{0}^{\alpha} \! -\delta/2, \: B_{0}^{\alpha} \! +\delta/2]$. $B_{0}^{\alpha}$ is the average value of the distribution and $\delta$ is a detuning parameter. 
The two--body residual interactions $J_{x}^{i,j}$ and residual system-bath couplings $\lambda^{i}_{\alpha}$ are also randomly and uniformly sampled within $[-J,\ J]$, and $[-\lambda,\ \lambda]$, respectively. Experimentally accessible control parameters for the charge-qubits\cite{Nori} are ${\cal{B}}^{\alpha}\!=1.00 \ {\epsilon}$, ${\cal{J}}_{x}=0.05 \ {\epsilon}$ and $kT=0.25\ {\epsilon}$, in units of $\epsilon=200$ mK. The total gate time is then $\tau_{9}=1.129\times 10^{-9}$ s. Since the idle and active qubits belong to the same QC, the idle qubits should only differ from active qubits by imperfections. Hence, the average of the distribution corresponds to $B_{0}^{\alpha}={\cal{B}}^{\alpha}$, and we set $\delta=0.4\ {\epsilon}$. We consider values $J_{x}= \: 0.05, \: 0.25, \: 0.50, \: 1.00, \: 2.00$, with units in ${\epsilon}$, to explore the onset of chaos and determine whether
the resulting bath-chaos is beneficial as predicted\cite{Prosen,Frahm,Tess,Mil}. Note that the spectrum is Wignerian above $J_{x}=0.15\ \epsilon$. (Exact diagonalization of the bath Hamiltonian is accomplished by a 
Lanczos algorithm\cite{ARPACK} for $N=10$ qubits.) 
 
In order to avoid error propagation due to initial residual entanglement between system and bath qubits, we assume an initial state of product form 
\begin{equation}
\hat{\rho} (0) = \hat{\rho}_{S} (0) \otimes \hat{\rho}_{B} (0)
\end{equation}
where $\hat{\rho}_{S}(0)$ is the density of the active system qubits and $\hat{\rho}_{B}(0)$ is the density of the idle bath qubits. In our simulations we consider eight different initial system states $\hat{\rho}_{S}(0)=|\psi_{0}\rangle\langle\psi_{0}|$ : four standard basis states 
\begin{equation}
|\psi_0\rangle \in \{|00\rangle,|01\rangle, |10\rangle, |11\rangle \},
\end{equation}
and four Bell states 
\begin{equation}
|\psi_0\rangle \in \{(|00\rangle \pm |11\rangle)/\sqrt{2}, (|01\rangle \pm |10\rangle)/\sqrt{2} \}.
\end{equation}
The assumption that the bath qubits are idle, and the fact that the residual two-body imperfections exist in the idle part, leads to dynamical thermalization of the bath qubits and hence the appropriate initial state is of canonical form. 

The reduced density of the system qubits is calculated as a partial trace over the bath degrees of freedom, 
\begin{equation}
\hat{\rho}_{S}(t)={\rm Tr}_{B}[\hat{U}(t)\hat{\rho}(0)\hat{U}^{\dagger}(t)]
\end{equation}
where the propagator is 
\begin{equation}
\hat{U}(t)=\hat{\cal T}\exp{[-(i/\hbar)\int_0^t\hat{H}(t')dt']}
\end{equation}
and $\hat{\cal T}$ denotes the time ordering operator. The reduced density at time $t$ can be expressed exactly as 
\begin{equation}
\hat{\rho}_{S}(t)=\sum_{n=1}^{n_{eig}} \frac{e^{-E_{n}/kT}}{Q}~{\rm Tr}_{B} 
[ |\Psi_{n}(t) \rangle \langle \Psi_{n}(t)| ]
\end{equation}
where the partition function is 
\begin{equation}
Q= \sum_{n=1}^{n_{eig}}e^{ -E_{n}/kT}
\end{equation}
and  
$
\hat{H}_{B} | \phi_{n}^{B} \rangle = E_{n} | \phi_{n}^{B} \rangle
$, and 
$|\Psi_{n}(t) \rangle$ is a solution of  the Schr\"{o}dinger equation
\begin{equation}
d|\Psi_n(t)\rangle/dt=-(i/\hbar)\hat{H}(t)|\Psi_n(t)\rangle
\end{equation} 
with initial condition 
$|\Psi_n(0)\rangle=|\psi_0\rangle\otimes |\phi_{n}^{B} \rangle$. Solutions of the Schr\"{o}dinger equations were obtained using a variable stepsize Runge-Kutta code\cite{RK}. The number of initial states which are thermally populated is small at low temperatures and so only a few need be propagated.

\section{Results}
Two measures of errors were employed to quantify errors during the CNOT gate: average purity, and average fidelity. For each we compute two averages: over the four initial standard basis states, and over the four initial Bell states. The average purity, 
\begin{equation}
{\bar{\mathcal P}}(t)=\frac{1}{4}\sum_{|\psi_{0}\rangle}{\rm Tr}_S [\hat{\rho}_{S}^{2}(t) ]
\end{equation}
measures proximity of the reduced density to a pure state. In the absence of system-bath coupling ${\bar{\mathcal P}}(t)=1$ at all times for pure initial states. The average fidelity, 
\begin{equation}
{\bar{\mathcal F}}(t) = \frac{1}{4}\sum_{|\psi_{0}\rangle}{\rm Tr}_S [\hat{\rho}_{S}(t)\hat{\rho}_{S}^{ideal}(t)]
\end{equation}
 measures how close the computed result is to the desired algorithm. In the absence of system-bath coupling, ${\bar{\mathcal F}}(t) = 1$, and ideal time evolution is given by 
 \begin{equation}
 \hat{\rho}_{S}^{ideal}(\tau_{9})=\hat{U}_{ {\rm CNOT} }\hat{\rho}_{S}(0)\hat{U}_{ {\rm CNOT} }^{\dagger}.
\end{equation}

\begin{figure}
\centering
\includegraphics[scale=0.35]{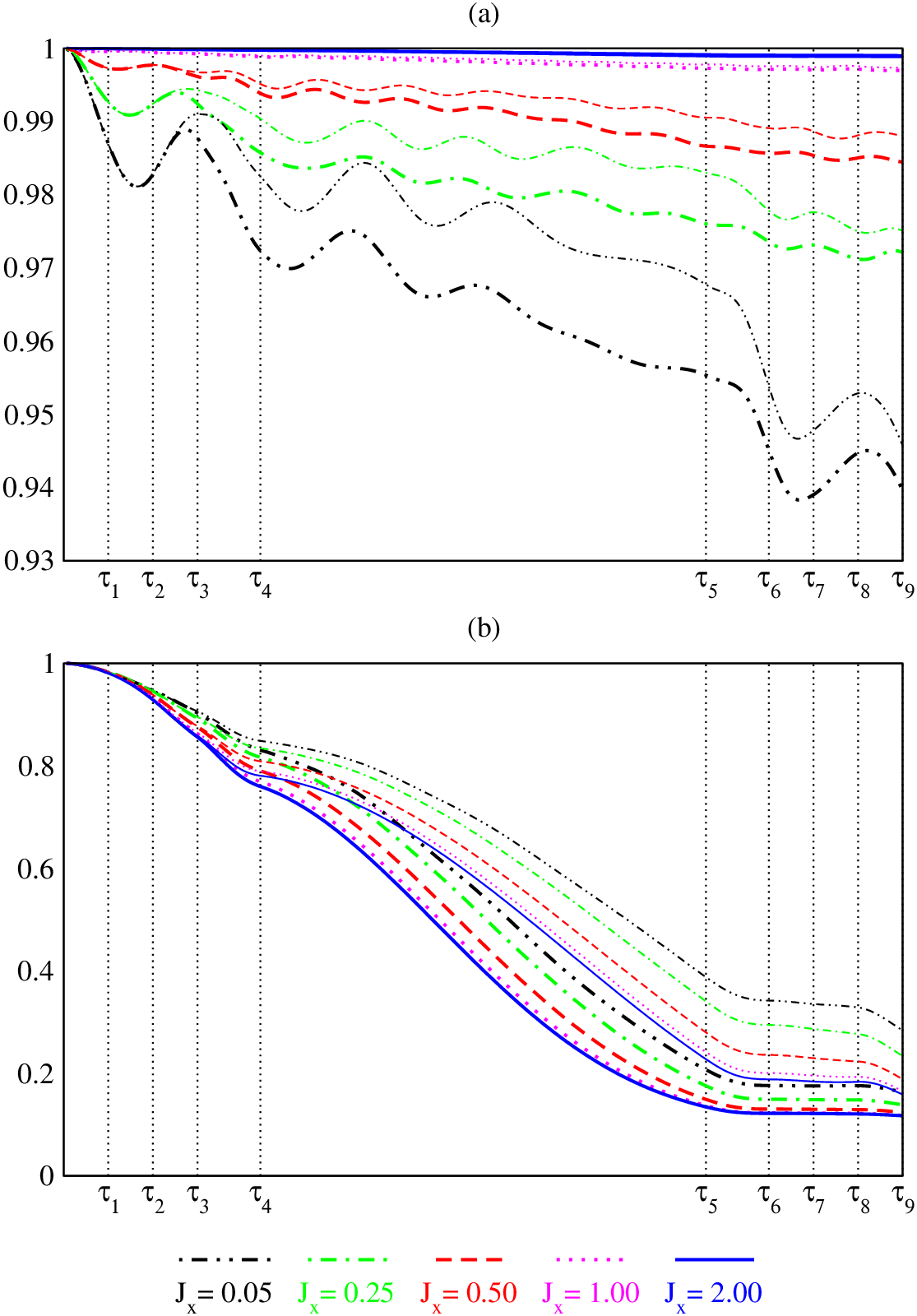} 
\caption{(Color online) (a) Average purity $\bar{\mathcal{P}}(t)$, and (b) fidelity $\bar{\mathcal{F}}(t)$ vs. time for bit--flip errors.}
\label{purity}
\end{figure}
\begin{figure}
\centering
\includegraphics[scale=0.35]{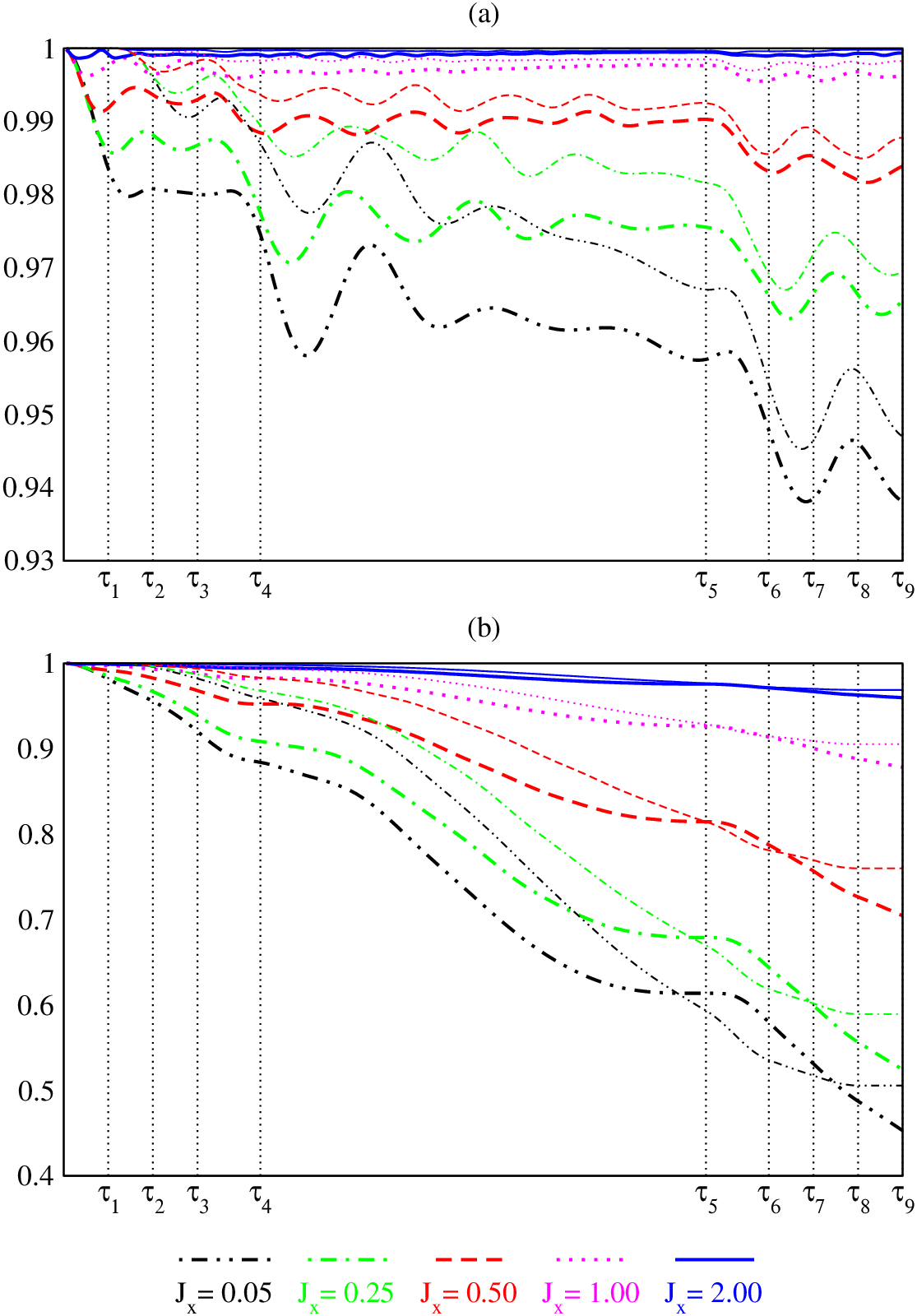} 
\caption{(Color online) (a) Average purity $\bar{\mathcal{P}}(t)$, and (b) fidelity $\bar{\mathcal{F}}(t)$ vs. time for phase errors.} 
\label{fidelity}
\end{figure}

In Fig. (1a) we plot average purity vs time, and in Fig. (1b) average fidelity vs time for bit-flip errors for five different values of intra-bath couplings $J_{x}$. Similar quantities are plotted in Fig. 2 for phase errors. In the Figures, the Bell states are distinguished from standard basis states by thick lines. 
Non-negligible deviations from perfect purity, greater than theoretically acceptable limit\cite{limit} of $0.99999$, are observed in almost all cases, indicating that internal decoherence exists even for this relatively small number of bath qubits. However, with increasing $J_{x}$, a transition to chaos occurs, which in turn results in rapid suppression of non-unitary errors. The strongest intra--bath coupling $J_{x}=2.00 \ \epsilon$ leads to nearly complete suppression of decoherence, which is true for both bit-flip and phase errors. Hence, deliberately induced bath chaos seems beneficial, and may even serve as error correction strategy when such strong interactions are practical. As expected, Bell states are more fragile to the effects of internal imperfections, and non-unitary bit-flip errors are of the same magnitude as phase-errors. 

The fidelity plots show a very large and unexpected deviation from ideality
even though the purity does not show this behavior. Hence, the large discrepancies between the purity and fidelity plots indicate that the environment primarily
induced unitary errors via coherent shifting processes. Increasing $J_{x}$ results in an increase in fidelity for phase errors, but no-improvement is observed  
for bit-flip errors. The surprising magnitude of the observed
unitary errors destroy practical QC operations, and methods
need to be devised for incrementally correcting the dynamics so that
multi-gate algorithms are viable. Note also that this error, here observed for a two-qubit subsystem, is of a collective nature and
hence common to the entire active part which in general may consist of many qubits. While existing error correction strategies  
could be employed to remove this shift\cite{Brown}, the number of single-body operations would have to greatly increased {\em for each
qubit}. Since the shift we observe is unitary and of a collective nature a much simpler collective error correction scheme may be possible.

We performed our simulations on ten different realizations of the QC. The results we report here are typical. However, we encountered
exceptions. Accidental near degeneracies in the low lying bath energies can occasionally occur with increasing $J_{x}$, causing increased 
decoherence. Architectures involving xy-type two-qubit imperfections actually favor this behavior. In all cases, the large magnitude of the 
coherent shift is universal. 

\section{Discussions}

\begin{figure}
\centering
\includegraphics[scale=0.3]{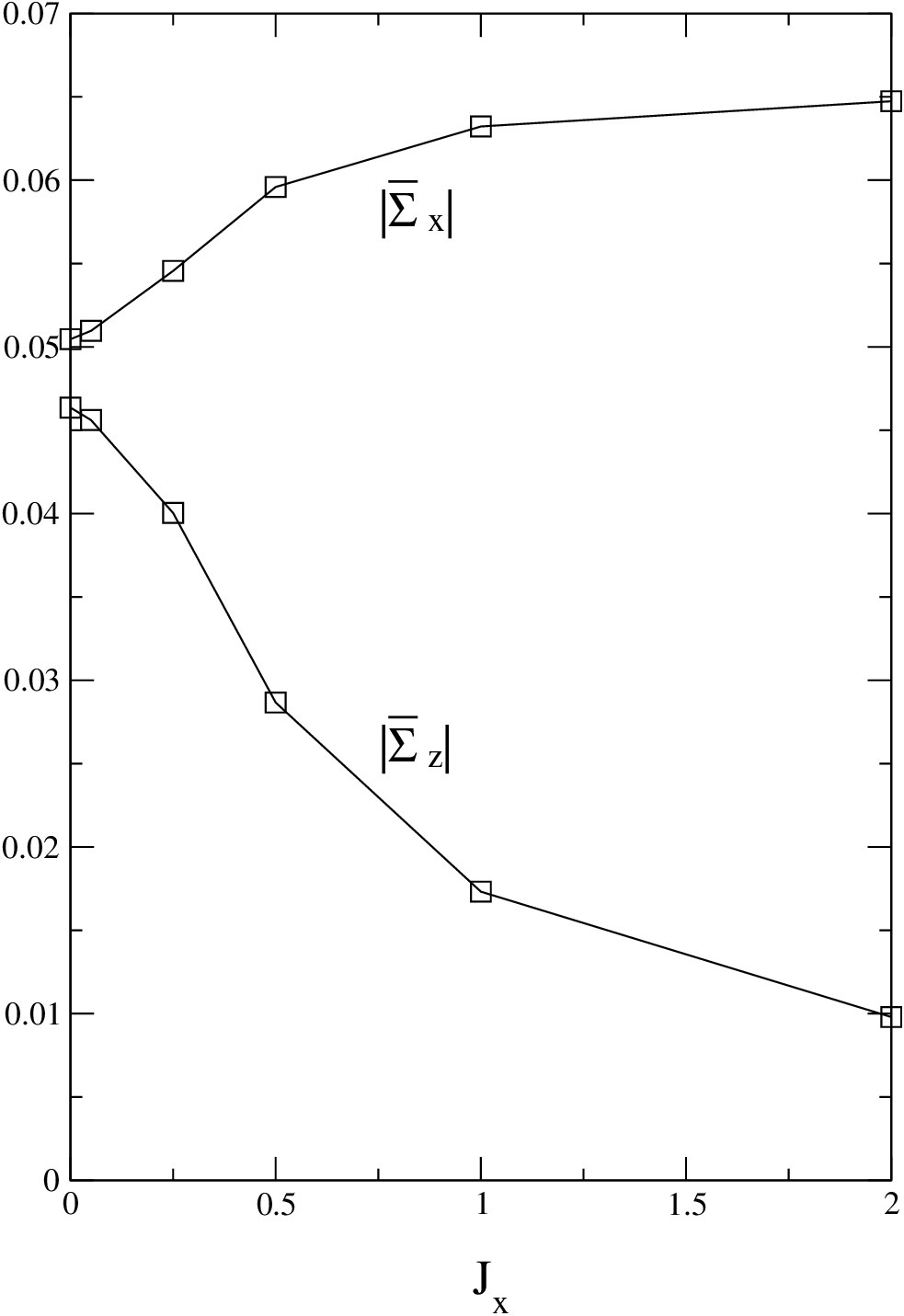} 
\end{figure}

The observed reduction of decoherence with increasing $J_{x}$ can be explained by  semiclassical arguments\cite{Tess,SemiC}. Basically, the squares of the off-diagonal matrix elements of the active-idle coupling operator - in the basis of bath eigenstates - scale as $\hbar^{N-1}$ when the bath is chaotic\cite{SemiC}. For small $\hbar$ they therefore vanish in
the thermodynamic limit. When this is combined with the fact that the diagonal elements vary slowly with energy in the low
energy region of the spectrum, it guarantees that decoherence and dissipation will be less than would be found in an integrable 
bath where selection rules are operative. 

The coherent shift is a well known feature of the exact Nakajima-Zwanzig master equation\cite{Zwan}, which is approximately of the form
\begin{equation}
\frac{d}{dt}\hat{\rho}_{S}(t)=-(i/\hbar)[\hat{H}_{S}(t)+\hat{S} \bar{\Sigma},\hat{\rho}_{S}(t)] +\hat{\cal L} \hat{\rho}_{S}(t) \label{shift}
\end{equation}
\noindent
in the Markovian limit. (See also our detailed discussion of a similar shift in \cite{CW2}.) Here $\hat{S}=(\hat{\sigma}_{\alpha}^{1}+\hat{\sigma}_{\alpha}^{2})$ is the system coupling operator and 
$\bar{\Sigma}={\rm Tr} [ \hat{\Sigma}_{\alpha}\hat{\rho}_{B}(0) ]$ is the canonical average for the bath coupling operator. $\hat{\cal{L}}$ 
includes non-unitary contributions to the dynamics. Clearly, the coherent shift originates from the dressing term $\hat{S} \bar{\Sigma}$ in Eq. (\ref{shift}). In Fig. 3 we plot $|\bar{\Sigma}_{x}|$ and $|\bar{\Sigma}_{z}|$ as a function of $J_{x}$. Increasing $J_{x}$ results in a small amount of increase in magnitude of $|\bar{\Sigma}_{x}|$, but a considerable amount of decrease in magnitude of $|\bar{\Sigma}_{z}|$, which is in very good agreement with the amount of decrease in fidelity for bit-flip errors in Fig. 1(b) and increase in fidelity for phase errors in Fig. 2(b). Chaos thus affects the shift indirectly 
through canonical averages which change with the bath eigenstates (i.e. $\bar{\Sigma}$ varies with $J_{x}$ through $|\phi_{n}^{B} \rangle$). Since bit-flip and phase error generators do not commute, each has a different  $\bar{\Sigma}$ 
even for the same bath states, and thus their behaviors differ. It should also be noted that since increasing bath 
chaos results in improved purity, a corresponding improvement in fidelity is expected. However, unitary errors due 
to the shift are so strong that fidelity improvement is not noticeable.    

The fact that the coherent shifting is sensitive to intra-bath coupling strength is suggestive. First, it means that the shifting
is at least partly a collective phenomenon. Second, while the shifting is clearly undesirable in a QC it may 
actually be useful in other situations. Note that Eq. (\ref{shift}) predicts that no shifting should be observed for a system coupled to a harmonic oscillator bath via a coordinate type coupling. In cases where shifting is observable it then gives 
a measure of anharmonicity and internal coupling of the bath degrees of freedom. Shifts observed in different types of
optically active impurities could thus be employed to obtain detailed information about a solid. 

\section{Conclusion}
In summary, the external decoherence time is not the only factor which should be taken into account in QC design and error correction strategies. 
Qubits in flawed QCs are subject to destructive effects of internal decoherence, dissipation, and coherent shifting. Bath-chaos may serve to correct non-unitary errors but the unitary errors due to the coherent shifting remain. While existing error correction codes may cure unitary errors, this will
greatly increase the number of required one-qubit rotations. We hope that our preliminary results promote further study of architecture
specific internal errors and strategies against them.

\ack{The authors acknowledge the support of the Natural Sciences and Engineering Research Council of Canada. Computations
were done using WestGrid computing resources.}

\end{document}